\documentclass[twocolumn,aps,prl,showpacs,superscriptaddress,longbibliography,10pt]{revtex4-1} %

\usepackage[colorlinks=true,linkcolor=blue,citecolor=blue]{hyperref}
\usepackage{amsfonts}
\usepackage{subfigure}
\usepackage{amsmath}
\usepackage{mathrsfs}
\usepackage{amssymb}
\usepackage{amsbsy}
\usepackage{epsfig}
\usepackage{graphicx}
\usepackage{epstopdf}
\usepackage{mathdots}
\usepackage{color}
\usepackage{cleveref}
\usepackage{float}
\usepackage{graphicx}
\usepackage{CJK}

\begin{document}
\begin{CJK*}{UTF8}{gbsn} 
\title{Topological energy braiding of the non-Bloch bands}

\author{Yang Li}
\affiliation{Department of Physics, Jiangsu University, Zhenjiang, 212013, China}

\author{Xiang Ji}
\affiliation{Department of Physics, Jiangsu University, Zhenjiang, 212013, China}

\author{Yuanping Chen} \altaffiliation{ypchen@ujs.edu.cn}
\affiliation{Department of Physics, Jiangsu University, Zhenjiang, 212013, China}

\author{Xiaohong Yan} 
\affiliation{Department of Physics, Jiangsu University, Zhenjiang, 212013, China}

\author{Xiaosen Yang} \altaffiliation{yangxs@ujs.edu.cn}
\affiliation{Department of Physics, Jiangsu University, Zhenjiang, 212013, China}

\date{\today}

\begin{abstract}
The non-Hermitian skin effect, as a unique feature of non-Hermitian systems, will break the topological energy braiding of the Bloch bands in open boundary systems. Going beyond the Bloch band theory, we unveil the energy braiding of the non-Bloch bands by introducing a one-dimensional non-Hermitian tight-binding model.
We find an entirely new generic class of topological non-Bloch bands such as Hopf link, which is generally generated by the non-Hermitian skin effect. The energy braiding is topologically robust against any perturbations without gap closing. Furthermore, non-Bloch topological invariants are proposed based on the generalized Brillouin zone to characterize the topology of these non-Bloch bands. The topological phase transition between the distinct phases occurs with the non-Bloch bands touching at exceptional points. We hope that our work can shed light on the topological energy braiding of the non-Bloch bands for non-Hermitian systems.
\end{abstract}

\maketitle
\end{CJK*}

Non-Hermitian (NH) Hamiltonians \cite{benderPhysRevLett1998, Bender2007, Rotter2009, huicaoRevModPhys2015, TonyPhysRevLett2016, zhongwangPhysRevLett2018a,  BergholtzRevModPhys2021}, applicable to a wide range of open systems such as photonic\cite{malzardPhysRevLett2015, PanNC2018, OzawaRevModPhys2019, Zhou2019Optic, XiaoLeiNaturePhysics2020, ZhuPhysRevResearch2020,XiaoLeiNaturePhysics2020, WangPhysRevLett2021}, cold atomic \cite{LeePhysRevX2014, XuPhysRevLett2017, WenPhysRevLett2019, hearxiv2020,zhaihuiNP2020,Qian2022NHC}, and acoustic \cite{WangPRL2018, ShenPhysRevMaterials2018, GaoNC2021, ZhangNC2021, ZhangNC2021a} systems with gain/loss, have attracted immense growing interest, especially in the topological properties \cite{SilveirinhaPhysRevB2019,  zhou2020topological,  LinhuPhysRevLett2020, WeiweiPhysRevLett2018, GuPhysRevApplied2021b}. In contrast to Hermitian systems, NH systems have complex eigenenergies, which give rise to unique phenomena without Hermitian counterparts like the exceptional points (EPs) \cite{Zhouscience2018, MiriScience2019, HahnNC2016, KawabataPhysRevLett2019, RuiPhysRevB2019, Yokomizo2020PhysRevResearch, ZhangPRB2020, XiaoPhysRevLett2021}, non-Hermitian skin effect (NHSE) \cite{OkumaPhysRevLett2020, zhongwangPhysRevLett2018a, PhysRevLettwangzhong2019a, LonghiPhysRevResearch2019, chenshuPRB2019, LeePhysRevLett2019, li2020critical, liu2020helical, HofmannPRR2020, yi2020nonhermitian, LiuShuoResearch2021, Guo2021PhysRevLett, HagaPhysRevLett2021,WangPhysRevLettBurst}, and non-Bloch bulk-boundary correspondence\cite{YokomizoPhysRevLett2019, KunstPhysRevLett2018, JinLPhysRevB2018, ghatak2019observation, haijunPhysRevB2019, EdvardssonPhysRevB2019, HelbigNaturePhysics2020, LonghiPhysRevLett2020, kouPhysRevB2020, Zhang2020PhysRevB, CaoPhysRevB2021, ZirnsteinPhysRevLett2021}. The NH systems can be described by generalizing the Brillouin zone(BZ) into a complex plane. For one-dimensional (1D) NH systems with a line or point gap \cite{OkumaPhysRevLett2020} in the complex energy plane, the topologically distinct phases can be classified by the non-Bloch topological invariants defined on the generalized Brillouin zone (GBZ)\cite{zhongwangPhysRevLett2018a, YokomizoPhysRevLett2019}, for example, non-Bloch winding numbers\cite{PhysRevLettwangzhong2019b, YokomizoPhysRevLett2019} and non-Bloch Chern numbers\cite{zhongwangPhysRevLett2018b, HirsbrunnerPRB2019, BrzezickiPRB2019}. 
NHSE and non-Bloch topological invariants are of fundamental importance for understanding of topological phenomena in non-Hermitian systems.

Recent works show that the complex eigenenergies of the NH systems can form intricate band structures \cite{ShenPRL2018, wojcikarxiv2021}. The Bloch bands of the NH periodic boundary systems can separate from each other in momentum space. Then, the complex energies of the separable Bloch bands can braid topologically and correspond to the conjugacy classes of the braid group\cite{Fannature2021knot, Hu2021PRLKnots, yuarxiv2021}.  For the periodicity of the BZ, the knot invariants completely classify the separable Bloch bands for 1D NH periodic boundary systems\cite{Kauffman2013Knots, WojcikPhysRevB2020}. However, these topological invariants defined on BZ can not characterize the topological properties of the NH open boundary systems. In the presence of NHSE, the complex eigenenergies will collapse, which completely devastated the topological energy braiding of Bloch bands in the momentum space. Therefore, is there any non-Bloch bands with topological energy braiding?  How to characterize the topological properties of the non-Bloch bands?

In this paper, we investigate the topological energy braiding in 1D NH open boundary systems. Going beyond the Bloch band theory, we find an entirely new generic class of topological non-Bloch bands with nontrivial energy braiding, which is generally generated by NHSE. The energies of the Bloch bands do not braid with each other along BZ under periodic boundary condition (PBC). But the one of the non-Bloch bands braid topologically along GBZ under open boundary condition (OBC), such as Hopf link. 
We propose the non-Bloch winding number and vorticity to characterize the topological energy braiding. The topological phase transition occurs when the gap closes at EPs. Furthermore, the energy braiding is topologically robust against any perturbation without gap closing. Our work paves the way to classify the NH topology for a wide range of 1D NH systems.

\begin{figure}
	\includegraphics[width=0.45\textwidth]{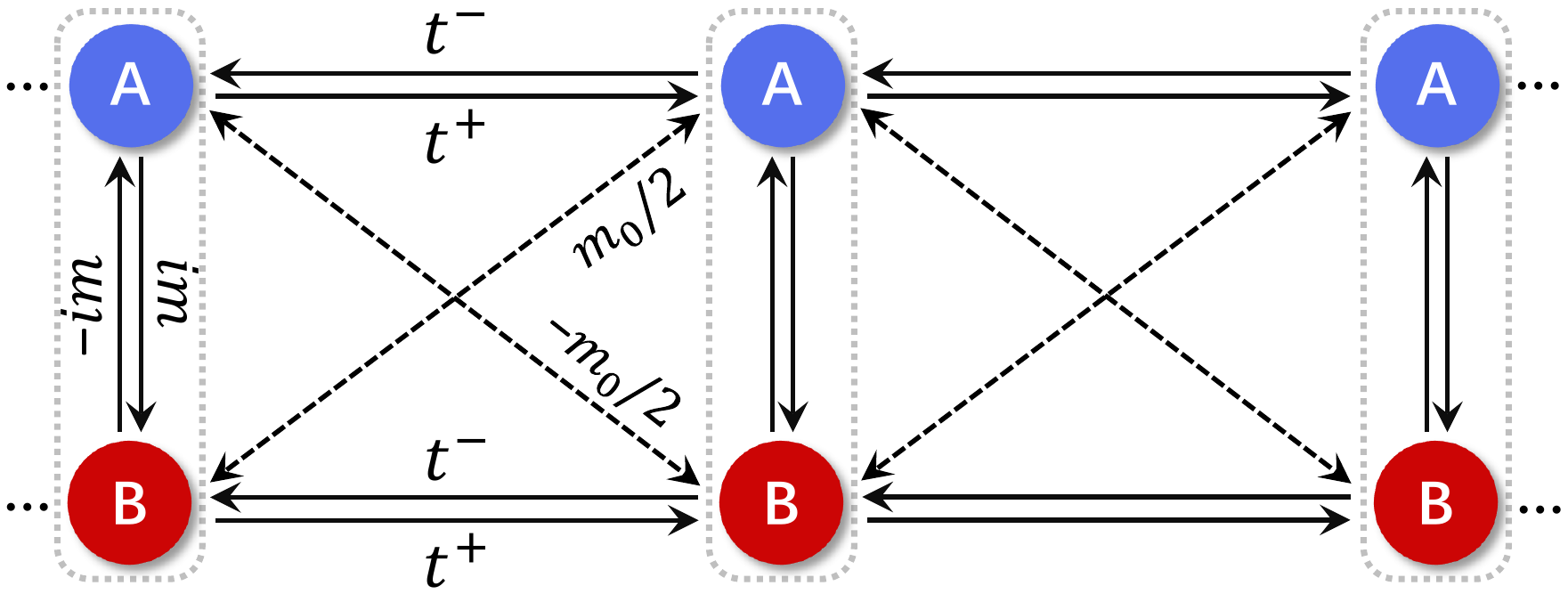}
	\caption{Sketch of the two bands non-Hermitian lattice model with non-reciprocal nearest-neighbour hopping $t^{\pm}$, which breaks the Hermicity of the system. The dotted box indicates a unit cell with two sublattices A and B.}\label{model}
\end{figure}

{\it Non-Hermitian model.--} To explore the complex-energy braiding and the topological properties of the non-Bloch bands, we construct a 1D tight-binding lattice model as shown in Fig.\ref{model}, whose Hamiltonian in the real space takes the form:
\begin{eqnarray}
&& H =\sum_{n} \sum_{\sigma\in \{A,B\}} \left(t^- c_{n,\sigma}^{\dag} c_{n+1,\sigma}+t^+ c_{n+1,\sigma}^{\dag} c_{n,\sigma}\right) \nonumber\\
&& \quad \quad  + \frac{m_0}{2}\left( c_{n+1,A}^{\dag}c_{n,B}-c_{n+1,B}^{\dag}c_{n,A}+H.c.\right) \nonumber\\
&& \quad \quad - \left( im c_{n,A}^{\dag}c_{n,B}+H.c. \right),
\label{Hamilt}
\end{eqnarray}
here $c^{\dag}_{n, \sigma} \, (c_{n, \sigma})$ is the creation (annihilation) operator of the sublattice $\sigma \in \{A,B\}$ on the $n$th site. $t^{\pm} = (t \pm \Delta)/2$ are the non-reciprocal nearest-neighbour hopping in the same sublattice. The Hermicity of the system is broken by the nonzero parameter $\Delta$. $m_0$ is the reciprocal nearest-neighbour hopping in different sublattices and $m$ is intracell coupling. The parameters $t$, $\Delta$, $m_{0}$ and $m$ are real.

\begin{figure*}
	\includegraphics[width=0.9\textwidth]{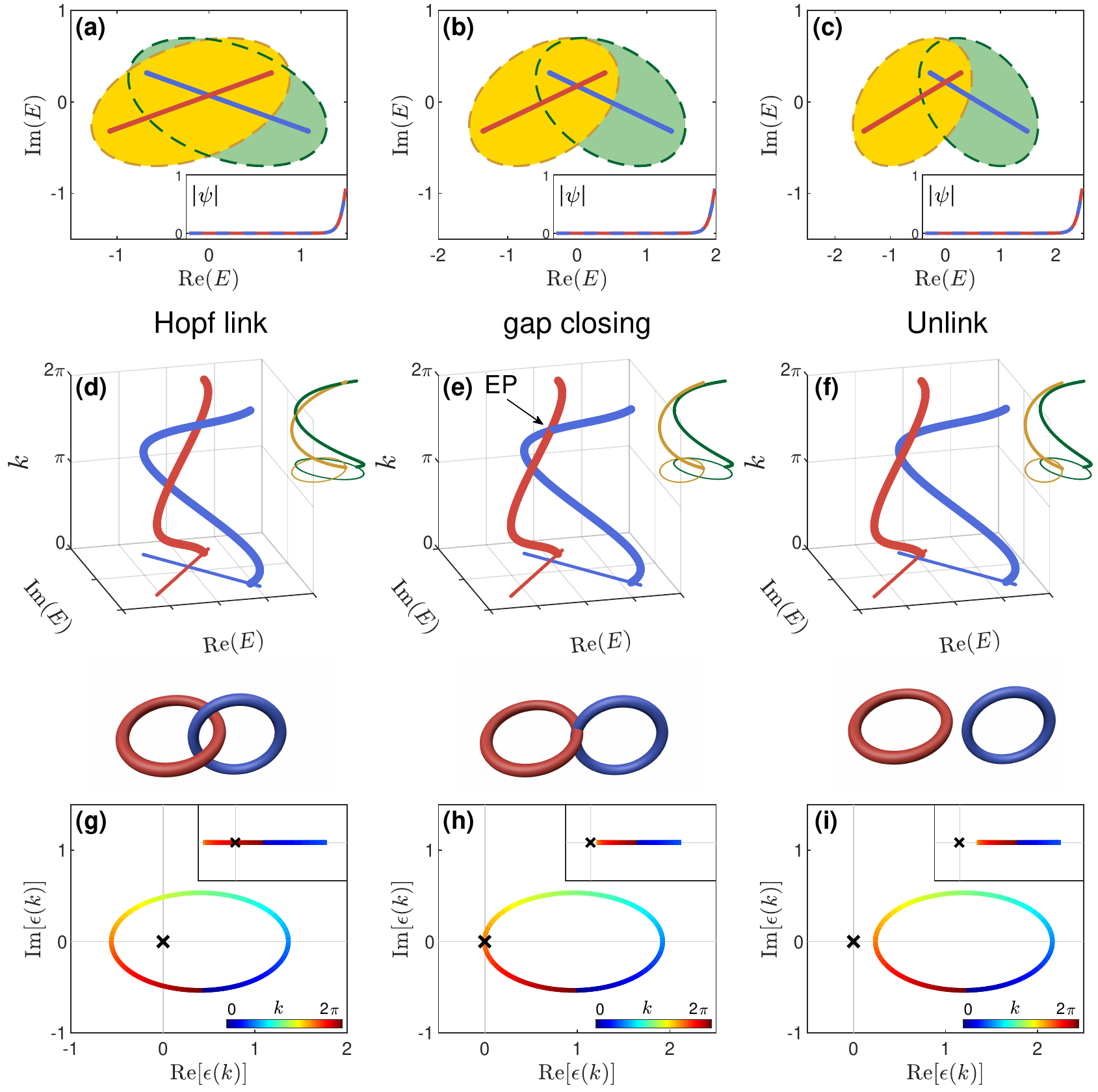}
	\caption{The complex-energies of the open lattice (solid curves) and periodic lattice (dashed loops) in the complex energy plane for (a) $m=0.2$, (b) $m=0.48$ and (c) $m=0.6$ with $t=1.0$, $\Delta=0.7$, $m_0=0.4$. The OBC energies enclose zero areas in contrast to the PBC energies, which enclose nonzero areas (yellow and green) in the complex energy plane. The inserts are the average of the eigenstates of the open boundary systems. The NH systems exhibit NHSE for the localization of the eigenstates. (d)-(f) The complex-energies braiding of the non-Bloch bands in the three-dimensional ($E,k$)-space. The separable non-Bloch bands of (d) and (f) are topologically distinct and correspond to Hopf link and unlink, respectively. (e) The critical case with two non-Bloch bands touching at an EP. The inserts show the Bloch bands without any topological energy braiding. (g) - (i) The separations of the non-Bloch bands along the GBZ. Topological phases transition occurs with separation closing $\epsilon(k_c)=0$ at EPs. The inserts are the separations of the Bloch bands. } \label{knot}
\end{figure*}

{\it Non-Hermitian skin effect and the energy braiding of non-Bloch bands.--}
For concreteness, we illustrate the NH system under PBC firstly. The Bloch Hamiltonian of the periodic  boundary system can be written on BZ as:
\begin{eqnarray}
H(k) = t \cos k -i \Delta \sin k + (m + m_0 \sin k) \sigma_{y}, \label{Bhamil}
\end{eqnarray}
where $\sigma_y$ is the Pauli matrix and $k$ is the real-valued Bloch wave vector. The complex eigenenergies
$E_{\pm}(k)=t \cos k-i \Delta \sin k \pm |m +m_0 \sin k| $ are given by the dashed loops in Fig.\ref{knot}(a) - Fig.\ref{knot}(c) for different $m$ with $\Delta=0.7$ and $m_{0}=0.4$. The PBC energies overlap each other and have a point gap in the complex energy plane.

For NH systems, the physical properties are sensitive to the boundary. The eigenenergies of the open NH system, denoted by the solid curves (red and blue), are different in essence from the PBC eigenenergies. Compared with the PBC eigenenergies, the OBC eigenenergies collapsed and enclosed zero areas in the complex energy plane. In addition, the OBC eigenstates localize at the right side of the open boundary system as shown in the inserts of Fig.\ref{knot}(a)-Fig.\ref{knot}(c). The NHSE and the collapsing of the OBC eigenenergies will certainly result in the non-Bloch band theory in characterizing the NH open boundary system  based on the GBZ\cite{zhongwangPhysRevLett2018a, YokomizoPhysRevLett2019}. The non-Bloch Hamiltonian $H(\beta)$ can be obtained from the Bloch Hamiltonian (Eq.\ref{Bhamil}) by extending the BZ into the GBZ with $\beta= e ^{i \tilde{k}}$ and complex-valued non-Bloch wave vector $\tilde{k} = k-i \ln r $. The eigenenergies of the non-Bloch bands take the form $E_{\pm}(\beta) = \pm m + 0.5(t-\Delta\mp im_0) \beta + 0.5 (t+\Delta\pm im_0)\beta^{-1}$.

In contrast to the Bloch band theory, the non-Bloch wave vector has an imaginary part that dictates the localization of the bulk eigenstates and induces the NHSE. The real part of non-Bloch wave vector $k$, nominated as momentum, is  restricted to a period $[0, 2\pi]$. For our system, the GBZ $C_{\beta}$ can be determined by the characteristic equation and can be written as $\beta= r e ^{i k}$  with the amplitude $r$:
\begin{equation}
r = \sqrt{\left|\frac{(t + \Delta + im_0)}{(t - \Delta - im_0)}\right|}.
\end{equation}

The non-Bloch bands, as described by $\beta(r,k)$-dependent complex energies, coincide with the eigenenergies of the NH open boundary system, which ensure a good definition of the non-Bloch band theory. Without loss of generality, we can investigate the non-Bloch band structures in the momentum $k$ instead of $\beta$. Furthermore, the complex energy of the non-Bloch bands and the momentum $k$ form a three-dimensional ($E,k$)-space. Then, the separable Bloch band structures of the periodic boundary systems\cite{ShenPRL2018} can be generalized into the open boundary systems based on the GBZ. We define a separation $\epsilon_{mn} (k) =  E_{m}(k) - E_{n}(k)$, which denotes the separation between the non-Bloch bands $m$ and $n$. A non-Bloch band is "isolated" if the energy does not overlap with that of any other non-Bloch band in the complex energy plane. A non-Bloch band $n$ is "separable" if the separation $\epsilon_{mn} (k) \neq 0$ for all $m \neq n$ and all $k$. The trajectory of the complex energy along the GBZ can be regarded as a strand of a braid in the three-dimensional $(E,k)$-space.

{\it Non-Bloch topological invariants.--} The ubiquitous twisting and braiding of the two separable non-Bloch bands can be classified by the braid group $B_{2}$. By generalizing the Bloch topological invariants\cite{ShenPRL2018,Fannature2021knot}, we propose a non-Bloch winding number for the non-Bloch bands on the GBZ as:
\begin{equation}
W = \frac{1}{2\pi i} \oint_{C_\beta} d \ln \det \left[ H(\beta)- \frac{1}{2} \text{Tr} \left(H(\beta) \right) \right].
\end{equation}

This non-Bloch winding number describes how many times the two  non-Bloch bands braid along the GBZ, with the sign indicating the handedness of the braiding. Two topologically distinct non-Bloch band structures cannot be continuously deformed into each other without the gap closing, where the non-Bloch winding number changes.

We show the complex eigenenergies of non-Bloch bands along the GBZ in Fig.\ref{knot}(d) - Fig.\ref{knot}(f). The two non-Bloch bands are separable in three-dimensional $(E,k)$-space, even though they overlap each other in the complex energy plane.  What's more remarkable, the separable non-Bloch band structures of Fig.\ref{knot}(d) and Fig.\ref{knot}(f) are topologically distinct. In Fig.\ref{knot}(d), the two non-Bloch bands braid around each other twice along the GBZ with the non-Bloch winding number $W=2$, which corresponds to a Hopf link phase. By contrast, the non-Bloch bands in Fig.\ref{knot}(f) do not braid around each other with non-Bloch winding number $W=0$ and correspond to unlink phase. There is a topological phase transition between topologically nontrivial Hopf link and trivial unlink phases at critical $m_c$, where the two non-Bloch bands touch at EPs as shown in Fig.\ref{knot}(e).
The inserts show the PBC energies in $(E,k)$-space. The PBC energies do not braid each other and have no topological band structures. Therefore, the topological Hopf link phase of the non-Bloch bands is indeed induced by the NHSE and the collapsing of the OBC energies. Then, we find that the collapsed OBC energies of the non-Bloch bands braid topologically in the momentum space.

\begin{figure}[H]
	\includegraphics[width=0.45\textwidth]{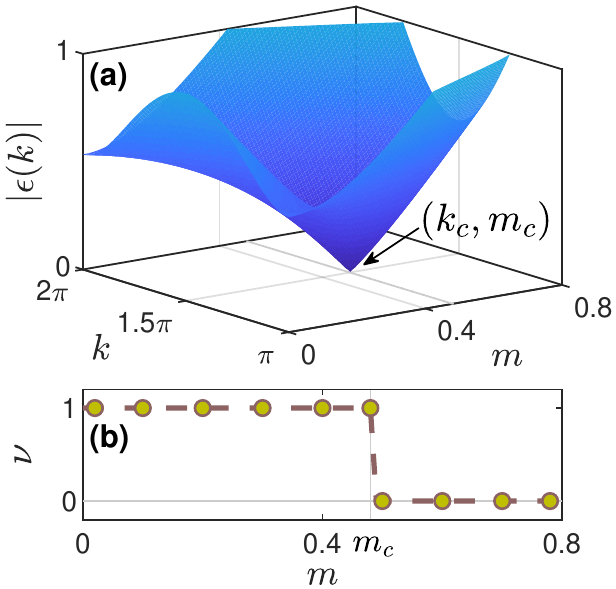}
	\caption{(a) The absolute value of the separation $\epsilon(k)$ of the non-Bloch bands as a function of $m$ and $k$ along the GBZs. The other parameters are identical to that of Fig.\ref{knot}. As increasing $m$, the separation has a node with $\epsilon(3\pi/2)=0$ at a critical value $m_{c}$, where the topological phase transition occurs. (b) The non-Bloch vorticity of the two non-Bloch bands as a function of $m$. The vorticity is $1$ for the topological Hopf link phase and $0$ for the unlink phase.}\label{separation}
\end{figure}

To visualize the topological phase transition, we show the separation $\epsilon (k)= \epsilon_{+-} (k)$ of the two non-Bloch bands in the complex energy plane along the GBZ in Fig.\ref{knot}(g) - Fig.\ref{knot}(i). The separation, always nonzero for the separable non-Bloch bands, encloses the EP for the topologically nontrivial Hopf link phase in Fig.\ref{knot}(g), but does not enclose it for the unlink phase in Fig.\ref{knot}(i). The topological phase transition between topologically distinct phases occurs only when the separation passes through the EP along the GBZ as shown in Fig.\ref{knot}(h). The inserts show the separations of the Bloch bands for periodic boundary systems, which do not enclose the EP. The two non-Bloch bands intersect at a critical momentum $k_{c}$ and $m=m_{c}$  with $\epsilon (k_{c})=0$ as shown in Fig.\ref{separation}(a). The intersection can be determined from the gap closing with $k_c = \frac{3\pi}{2}$ and $m_{c}= m_0(r^2 +1)/2r$ for $m>0$. For $m<0$ case, the separation is zero at $k_{c}=\pi/2$ and $m_{c}= -m_0(r^2 +1)/2r$.

The argument of the separation on the GBZ contributes to the topological invariants - non-Bloch vorticity:
\begin{equation}
\nu= \frac{1}{2\pi} \oint_{C_\beta} d \arg\left[\epsilon (\beta) \right].
\end{equation}

Fig.\ref{separation}(b) shows the non-Bloch vorticity varies with $m$. The vorticity, which is half of the non-Bloch winding number with $\nu= W/2$, is $1$ for the topological Hopf link phase and $0$ for the topologically trivial unlink phase. It changes an integer at the topological phase transition between the Hopf link and the unlink phases.

\begin{figure}
	\includegraphics[width=0.45\textwidth]{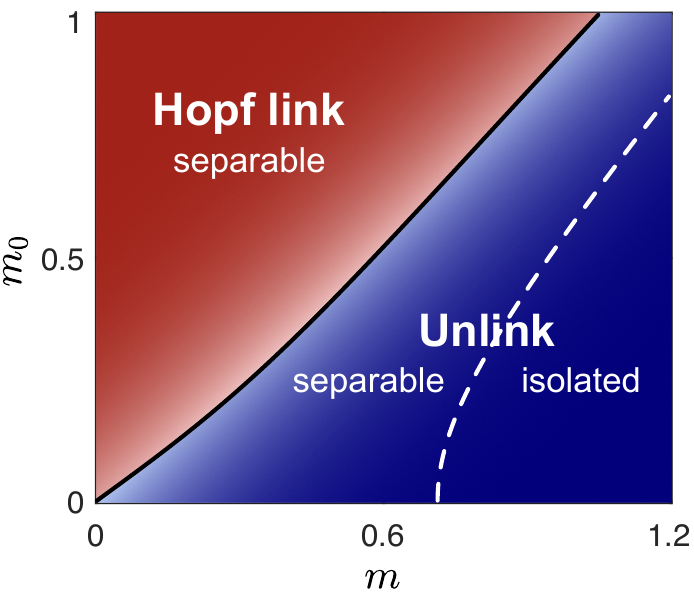}
	\caption{ Phase diagram of the non-Hermitian open boundary system in the $m-m_{0}$ plane with $t=1.0$ and $\Delta=0.7$. The dotted curve denotes the boundary between the phases of separable and isolated. For the separable regions, the solid curve is the boundary between topologically nontrivial Hopf link and the trivial unlink phases.}\label{OBCPD}
\end{figure}

{\it Phase diagram.--} In Fig.\ref{OBCPD} we show a phase diagram with respect to the parameters $m$ and $m_{0}$. The boundary between the separable and isolated phases can be determined with $m = \sqrt{ [(t-\Delta)r + (t+\Delta)/r]^2 + (m_0 r + m_0/r)^2 }/2 $. The separable phase is Hopf link with $m < m_0(r^2 +1)/2r$ and unlink for other regions. There is an intrinsic difference between phase diagrams of the open and the periodic boundary systems. For periodic boundary system, all the phases are topologically trivial without any nontrivial complex energy braiding. The topological energy braiding of the open boundary system is generated by the NHSE. In addition, the energy braiding of the non-Bloch bands is topologically robust against any perturbation without gap closing. For example, we can introduce a mismatched perturbation $\delta$ on the intracell coupling $m$. Although the GBZ is no longer a circle, the energies of the non-Bloch bands also topologically braid for the Hopf link phase. The perturbation only slightly changes the phase transition point without breaking the energy braiding. The robust properties of the topological energy braiding in the open boundary NH systems ensure the realization in future experiments.

{\it Conclusion.--} We have uncovered the topological energy braiding of the non-Bloch bands in a 1D NH open boundary systems.  Due to the NHSE, the non-Bloch band structures are topologically different from that of the Bloch bands. In particular, the complex energies of non-Bloch bands braid topologically in momentum space and correspond to the Hopf link, in contrast to the topologically trivial Bloch bands. Furthermore, we have proposed the non-Bloch winding number and vorticity to characterize the topological energy braiding based on the GBZ. The topological distinct phases can not be continuously transformed into each other without the gap closing at exceptional points. The energy braiding of the non-Bloch bands is topologically robust against any perturbation without gap closing. Our work can be extended to investigate other sophisticated knots and links in non-Bloch bands and high dimensionl NH systems, the effects of the non-Bloch PT symmetry, and so on. We anticipate that our work can pave the way to explore the topological energy braiding of the non-Bloch bands in NH systems.

{\it Acknowledgements.--} We thank Z. Wang for fruitful discussions. This work was supported by the National Natural Science Foundation of China (NSFC) under Grant 12174157 and 12074150.

\appendix

\section{Derivation of the GBZ $C_\beta$}

According to the Hamiltonian Eq.(1) in the main text, the real space eigen-equation leads to
\begin{equation}
\begin{split}
-im \psi_{n, B} + t^- \psi_{n+1, A} + t^+ \psi_{n-1, A} \\ - \frac{m_0}{2} \psi_{n+1, B}  + \frac{m_0}{2} \psi_{n-1, B} = E \psi_{n, A}  \\
im \psi_{n, A} + t^- \psi_{n+1, B} + t^+ \psi_{n-1, B} \\ + \frac{m_0}{2} \psi_{n+1, A}   - \frac{m_0}{2} \psi_{n-1, A} = E \psi_{n, B}  \\
\end{split}
\end{equation}

Taking the ansatz that $|\psi \rangle = \sum_{j}|\phi^{(j)}\rangle$, where $j$ is an independent component index of the wave function. Firstly, omitting the $j$ index and each $|\psi^{j} \rangle$ takes the exponential form: $\left( \phi_{n, A}, \phi_{n, B} \right) = \beta^{n} \left( \phi_{A}, \phi_{B} \right)$, which satisfies
\begin{equation}
\begin{split}
\left( -im - \frac{m_0}{2}\beta + \frac{m_0}{2}\beta^{-1} \right)\phi_B = \left( E - t^{-}\beta - t^{+}\beta^{-1} \right)\phi_A  \\
\left( im + \frac{m_0}{2}\beta - \frac{m_0}{2}\beta^{-1} \right)\phi_A = \left( E - t^{-}\beta - t^{+}\beta^{-1} \right)\phi_B
\end{split}
\end{equation}
Then, we have
\begin{eqnarray}
\phi_B = \pm i \phi_A
\end{eqnarray}
The eigen-equation in both above two cases are a quadratic equation for $\beta$. When $\phi_B = i \phi_A$, one can obtain
\begin{eqnarray}
\beta_1 \beta_2 = \frac{t+ \Delta +im_0}{t-\Delta -im_0}
\end{eqnarray}
while
\begin{eqnarray}
\beta'_1 \beta'_2 = \frac{t+ \Delta -im_0}{t-\Delta +im_0}
\end{eqnarray}
for $\phi_B = -i \phi_A$. The presence of two roots implies that the index of the wave function is $j=1, 2$. Therefore, restoring the $j$ index, the general solution is written as a linear combination:
\begin{equation}
\begin{split}
\psi_{n, A} = \beta_1^n \phi^{(1)}_A + \beta_2^{n} \phi^{(2)}_A  \\
\psi_{n, B} = \beta_1^n \phi^{(1)}_B + \beta_2^{n} \phi^{(2)}_B
\end{split}
\end{equation}
and $ \psi_{n, B} = \pm i \psi_{n, A} $, which should satisfy the boundary condition:
\begin{equation}
\begin{split}
-im \psi_{1,B} + t^- \psi_{2,A} - \frac{m_0}{2}\psi_{2,B} = E \psi_{1,A} \\
im \psi_{1,A} + t^- \psi_{2,B} + \frac{m_0}{2}\psi_{2,A} = E \psi_{1,B}  \\
-im \psi_{L,B} + t^+ \psi_{L-1,A} + \frac{m_0}{2}\psi_{L-1,B} = E \psi_{L,A}   \\
im \psi_{L,A} + t^+ \psi_{L-1,B} - \frac{m_0}{2}\psi_{L-1,A} = E \psi_{L,B}
\end{split}
\end{equation}
For the case of $\psi_{n, B} = i \psi_{n, A}$, the boundary condition is reduced to
\begin{equation}
\begin{split}
\left(m-E \right)\psi_{1,A} + \left( t^- - \frac{im_0}{2} \right)\psi_{2,A} =0  \\
\left(m-E \right)\psi_{L,A} + \left( t^+ + \frac{im_0}{2} \right)\psi_{L-1,A} =0
\end{split}
\end{equation}
Bringing it into the expression of the wave function to eliminate the unknown, we have
\begin{eqnarray}
\beta_1^{L+1} = \beta_2^{L+1}
\end{eqnarray}
which means $|\beta_1| = |\beta_2|$. Combined with $\beta_1 \beta_2 = (t+ \Delta +im_0)(t-\Delta -im_0)$, $|\beta_1| = |\beta_2|$ leads to
\begin{eqnarray}
|\beta_j| = r_+ = \sqrt{ \left| \frac{t+\Delta+im_0}{t-\Delta-im_0} \right| }
\end{eqnarray}
By the same method, when $\psi_{n, B} = -i \psi_{n, A}$, we have
\begin{eqnarray}
|\beta'_j| = r_- = \sqrt{ \left| \frac{t+\Delta-im_0}{t-\Delta+im_0} \right| }
\end{eqnarray}

\section{Non-Bloch winding number and the non-Bloch vorticity}

The non-Bloch Hamiltonian in terms of $\beta$ can be written as:
\begin{equation}
H(\beta) =
\left[
\begin{array}{cc}
t^- \beta + t^+ \beta^{-1}  &  -im-\frac{m_0}{2}\left( \beta-\beta^{-1} \right) \\
im+\frac{m_0}{2}\left(\beta-\beta^{-1}\right)  &  t^- \beta + t^+ \beta^{-1}\\
\end{array}
\right],
\end{equation}
with the eigenenergies
\begin{align}
E_{+}(\beta) &= m + \frac{t-\Delta- im_0}{2} \beta + \frac{t+\Delta+ im_0}{2} \beta^{-1},\\
E_{-}(\beta) &= -m + \frac{t-\Delta+ im_0}{2} \beta + \frac{t+\Delta- im_0}{2} \beta^{-1}.
\end{align}

From the non-Bloch Hamiltonian, the non-Bloch winding number can be simplified:
\begin{align}
W &= \frac{1}{2\pi i} \oint_{C_\beta} d \ln \det \left[ H_{\beta}- \frac{1}{2} Tr \left( H_{\beta} \right) \right] \nonumber\\
&= \frac{1}{2\pi i} \oint_{C_\beta} d \ln \left\{ \left[ E_+(\beta)- \frac{1}{2} Tr\left( H_{\beta} \right)\right] \left[ E_-(\beta)- \frac{1}{2} Tr\left( H_{\beta} \right)\right] \right\} \nonumber\\
&= \frac{1}{2\pi i} \oint_{C_\beta} d \ln \left\{ \frac{i}{2}\left[ E_+(\beta)-E_-(\beta) \right] \right\}^2 \nonumber\\
&= \frac{1}{\pi i} \oint_{C_\beta} d \ln \left\{ \frac{i}{2}\left| E_+(\beta)-E_-(\beta) \right| e^{ \arg\left[ E_+(\beta)-E_-(\beta) \right] } \right\} \nonumber\\
&= \frac{1}{\pi} \oint_{C_\beta} d \arg\left[ E_+(\beta)-E_-(\beta) \right] \nonumber\\
&= 2 \nu.
\end{align}
Then, we have the relationship between the two non-Bloch topological invariants: $W=2\nu$.

\section{The robust energies braiding of non-Bloch bands}
Here, we show that the energies braiding of the non-Bloch bands is topologically robust against perturbations for example at intracell coupling $m$:
\begin{eqnarray}
&& H =\sum_{n} \sum_{\sigma\in \{A,B\}} \left(t^- c_{n,\sigma}^{\dag} c_{n+1,\sigma}+t^+ c_{n+1,\sigma}^{\dag} c_{n,\sigma}\right) \nonumber\\
&& \quad \quad  + \frac{m_0}{2}\left( c_{n+1,A}^{\dag}c_{n,B}-c_{n+1,B}^{\dag}c_{n,A}+H.c.\right) \nonumber\\
&& \quad \quad - im c_{n,A}^{\dag}c_{n,B}+ (im+\delta) c_{n,B}^{\dag}c_{n,A}.
\label{Hamilt1}
\end{eqnarray}

\begin{figure}
	\includegraphics[width=0.45\textwidth]{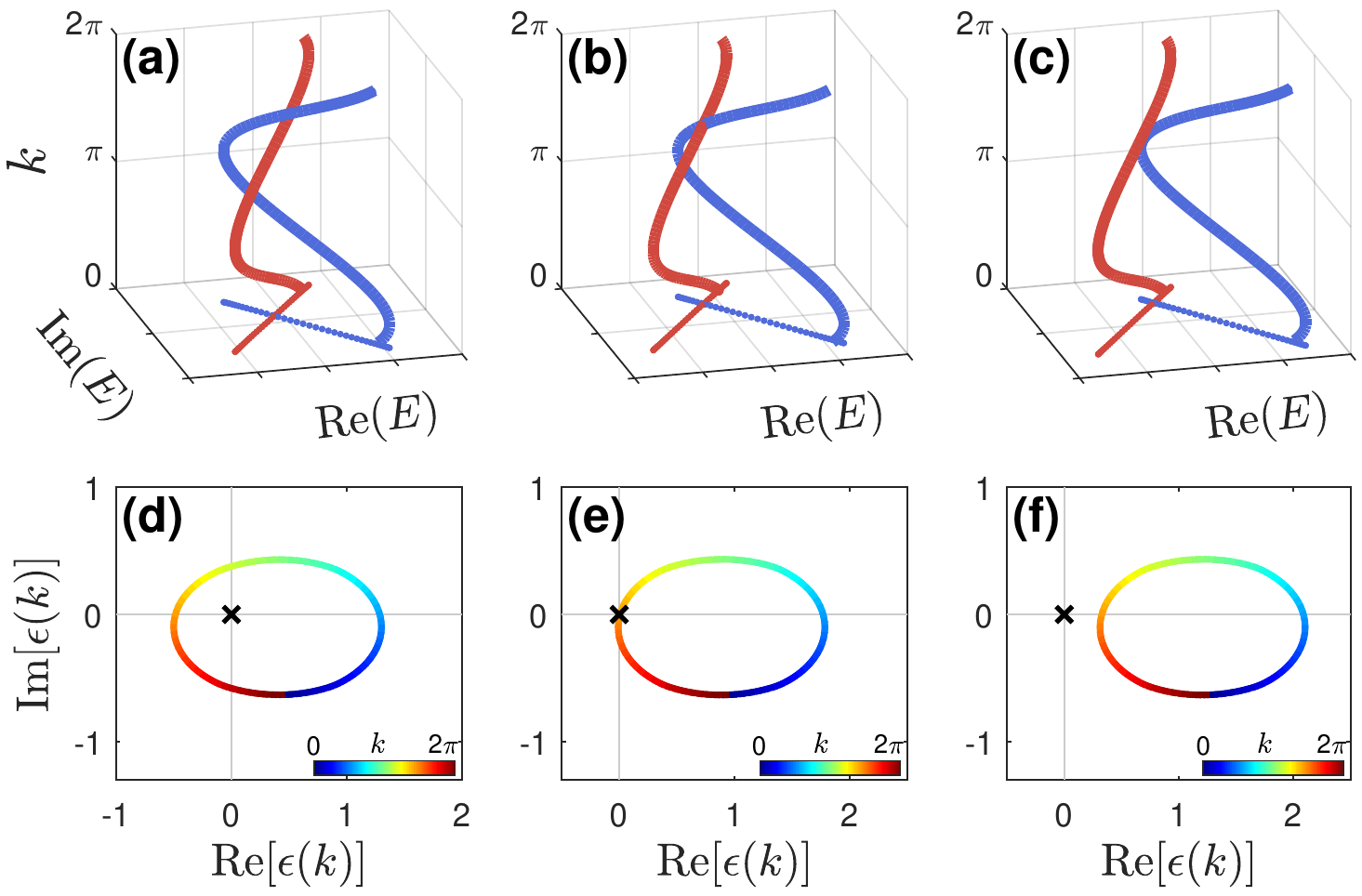}
	\caption{The complex-energies braiding of the non-Bloch bands in the three dimensional ($E,k$)-space for (a) $m=0.2$, (b) $m=0.46$ and (c) $m=0.6$ with $t=1$, $\Delta=0.7$, $m_0 = 0.4$ and $\delta=0.1$. The complex energies of the separable non-Bloch bands braid each other along GBZ and robust against $\delta$. (d)-(f) The separations of the non-Bloch bands along the GBZ.} \label{robust}
\end{figure}

The complex energies braiding of the non-Bloch bands are shown in Fig.\ref{robust}(a)-Fig.\ref{robust}(c) with $\delta=0.1$. The GBZ is no longer a circle. In Fig.\ref{robust}(a), the complex energies of the non-Bloch bands braid around each other twice along the GBZ, which corresponds to a Hopf link phase. The complex energies in Fig.\ref{robust}(c) do not braid around each other and correspond to unlink phase. There is a topological phase transition between Hopf link and unlink phases at critical $m=0.46$, where the two non-Bloch bands touch at EPs as shown in Fig.\ref{robust}(b). The separations are shown in Fig.\ref{robust}(d)-Fig.\ref{robust}(e). Therefore, the perburbation $\delta$ only changes the phase transition point but not breaks the topological energies braiding of the non-Bloch bands.

\bibliography{reference} 

\end{document}